\documentclass[journal=jacsat,manuscript=article]{achemso}
\usepackage[version=3]{mhchem} % Formula subscripts using \ce{}
\usepackage{siunitx}
\usepackage{subcaption}

\author{Jeet Majumdar}
\altaffiliation{Equal contribution}
\author{Subhadeep Dasgupta}
\altaffiliation{Equal contribution}
\author{Soham Mandal}
\author{Mohd Moid}
\author{Manish Jain}
\author{Prabal K. Maiti}
\email{maiti@iisc.ac.in}
\affiliation[Indian Institute of Science]{Centre for Condensed Matter Theory, Department of Physics, Indian Institute of Science, Bangalore 560012, India}

\title[An \textsf{achemso} demo]
  {Does twist angle affect the properties of  water confined inside twisted bilayer graphene?}

\abbreviations{monolayer water, nanoconfinement, graphene bilayer, molecular dynamics, structure, dynamics, dielectric, friction}

\begin{document}

\begin{abstract}
Graphene nanoslit pore is used for nanofluidic devices like water desalination, ion-selective channels, ionic transistors, sensing, molecular sieving, blue energy harvesting, and protein sequencing.
It is a strenuous task to prepare nanofluidic devices because a small misalignment leads to a significant alteration in various properties of the devices.
Here we focus on the rotational misalignment between two parallel graphene sheets.
Using molecular dynamics simulation, we probe the structure and dynamics of monolayer water confined inside graphene nanochannels for a range of commensurate twist angles.
With SPC/E and TIP4P/2005 water model, our simulations reveal the independence of equilibrium number density $(n \sim \SI{13}{\per\nano\meter\squared})$ for SPC/E and $(n \sim \SI{11.5}{\per\nano\meter\squared})$ for TIP4P/2005) across twists. 
Based on the respective densities of water models, the structure and dielectric constant are invariant of twist angles.
The confined water structure at this shows square ice ordering for SPC/E water only. 
TIP4P/2005 shows ordering at the vicinity of a critical density $(n \sim \SI{12.5}{\per\nano\meter\squared})$.
The average perpendicular dielectric constant of the confined water remains anomalously low ($\sim 2$ for SPC/E and $\sim 6$ for TIP4P/2005) for studied twist angles.
We find that the friction coefficient of confined water molecules varies for small twist angles while becoming independent for twists greater than $\SI{5.1}{\degree}$.
Our results indicate that small angular misalignment will not impair the dielectric properties of monolayer water within graphene slit-pore but can significantly influence its dynamics.
\end{abstract}

\section{I. Introduction}

Carbon based nanofluidic devices (CBND) are used in the emerging field of nanofluidics where one studies the fluid behaviour under various internal or external perturbations.
One of the well known CBND is graphene nano slit pore\cite{liu2021controlling} which has been used in molecular sieving\cite{sun2019selective, schlichting2020selective}, water desalination\cite{liu2021controlling, kurupath2021highly, sathian2020effect}, ionic transistor\cite{girdhar2013graphene, qiu2014graphene}, protein sequencing\cite{wasfi2020dna}, blue energy harvesting\cite{xie2022two}.
The fabrication of such devices based on graphene slit pore is an extremely delicate process because a small alteration can lead to significant changes in various properties of the desired final attributes.

The properties of confined water strongly depends on dimensionality\cite{moid2021dimensionality}, geometry of confinement \cite{borah2012transport, moid2019microscopic}, surface chemistry of the confining walls \cite{sam2019water, ayappa2019enhancing, kumar2018phase, ayappa2019enhancing} and its curvature \cite{marti2017structure, haji2017computational}.
The determinant of surface properties, such as hydrophobicity or hydrophilicity, is found to be contained in the microscopic properties of interfacial water \cite{bonthuis2012profile, sato2018hydrophobic, ayappa2020dynamical, rajasekaran2020influence}. 
An example of a complex surface water behaviour could be a change of water flux at the interfaces of CBND that can induce an oscillatory current across the carbon flake. 
This emerges from peculiar momentum transfer between charge carriers of the carbon walls and fluid molecules that are mediated through phonon excitations \cite{marcotte2022strong}, leading to anomalous behaviour in the dielectric constant of confined water. % MOID
Water closer to the surface exhibit reduced dipole moment fluctuations, partly due to the formation of ordered structures in some cases and partly due to the caging effect of water–surface interactions \cite{mosaddeghi2012simulations, mondal2020water}.
The first experimentally observed room temperature square ice formation was reported by Algara-Siller \emph{et al.} \cite{algara2015observation} using transmission electron microscopy, where they showed how a monolayer of water sandwiched between two graphene sheets, takes the form of a square ice with a lattice constant of about $\SI{2.8}{\angstrom}$.
Following this experiment, several computational works have been done to study the properties of this square ice using molecular dynamics (MD) simulations, employing a variety of water models.\cite{majumdar2021dielectric, corsetti2016structural}
However, this square ice ordering inside graphene pore is still highly debatable, as first principle calculation shows hexagonal-close-packed structure at low pressure regime at  300 K \cite{chen2017double, chen2016evidence, chen2016two}.
Indeed, the experimental observation by Algara-Siller \emph{et al.} was initially questioned with the possibility of sample contamination during experiment\cite{zhou2015observation}. 
We did not find any other similar experimental observation of confined water under identical physicochemical conditions.
Meanwhile, in subsequent simulation works, this ordering at ambient conditions was only observed for a certain group of water models which is also highly sensitive to lateral pressure and slit-width \cite{dix2018different, majumdar2021dielectric, zhu2016ab}.
Thus, highlighting the anomalous nature of confined water and the experimental challenges involved in reaching a conclusion on the formation of square ordering of confined water.

The static perpendicular dielectric $(\epsilon_{\perp})$ constant of confined water near the wall of graphene slit-pore also highlights this anomaly\cite{fumagalli2018anomalously,bonthuis2012profile, jalali2020out}.
Along with the static dielectric property, the friction coefficient too shows a peculiar behaviour. 
To this end, there have been multiple attempts to understand the dependence of friction that might originate due to periodic potential at nano scale\cite{xue2022peeling}, hydro-electronic coupling\cite{marcotte2022strong} and internal vibrations\cite{zhou2018kinetic}.
However, at the atomic and molecular scale description, the water friction was found to be dependent on atomic interaction with different solid surfaces\cite{tocci2014friction, goswami2020exploring, metya2018ice}, wetting properties\cite{cottin2003low}, lattice constant\cite{yang2020dynamics}, defects\cite{seal2021modulating}, and geometry and height of confinement\cite{yang2020dynamics, kavokine2022fluctuation, lee2017enhancement, senapati2001dielectric, babu2012combining, suraj2021salt}.
And although the strong $\sigma-\text{bonds}$ of in-plane carbon atoms and weak $\pi-\text{bonded}$ interactions between the atoms of adjacent layers allows graphite to slip on each other easily, intercalated water within these carbon layers can further enhance their lubrication and modify surface properties.
The diffusivity of the surface water molecules approaches to the bulk as one recedes away from the surface \cite{zhao2020molecular}. 
Thus, graphite lubrication ensue its tribological property from the microscopic interaction between water layer that is present between the hexagonal planes of graphene.
In this context, monolayer of water confined inside slit-pore channel of bilayer graphene can serve as an excellent model system to understand the properties of interfacial water in reduced dimension.
From an experimental perspective, where fine control of graphene layers is still challenging, the behaviour of confined water inside twisted bilayer graphene (tBLG) for various twist angles of the confining graphene sheets is of particular importance.
In this work, we study the effect of twist angles of tBLG on the structure, permittivity, and friction coefficient of monolayer water confined inside infinite slit-pore geometry using classical all-atom MD simulations.
We examined two classes of water models SPC/E and TIP4P/2005 inside our confined geometry. 
SPC/E - a model which is one of the widely used water model that often serves as solvent, and TIP4P/2005 - whose ordering is closest to the results obtained from first principle calculation of water under such extreme confinement.
We demonstrate that these properties do not show any significant change with varying twist angle of bilayer graphene.

The rest of the article is organized as follows: in section II \ref{modelling_methods}, we describe about the model and method. 
Section III describes the results for the structure of a monolayer of water molecules confined between twisted bilayer graphene sheets. 
The final section contains a summary and discussion of the main results.

\section{II. Modelling and methods}
\label{modelling_methods}
\begin{figure}[ht]
    \centering
    \includegraphics[width=\textwidth]{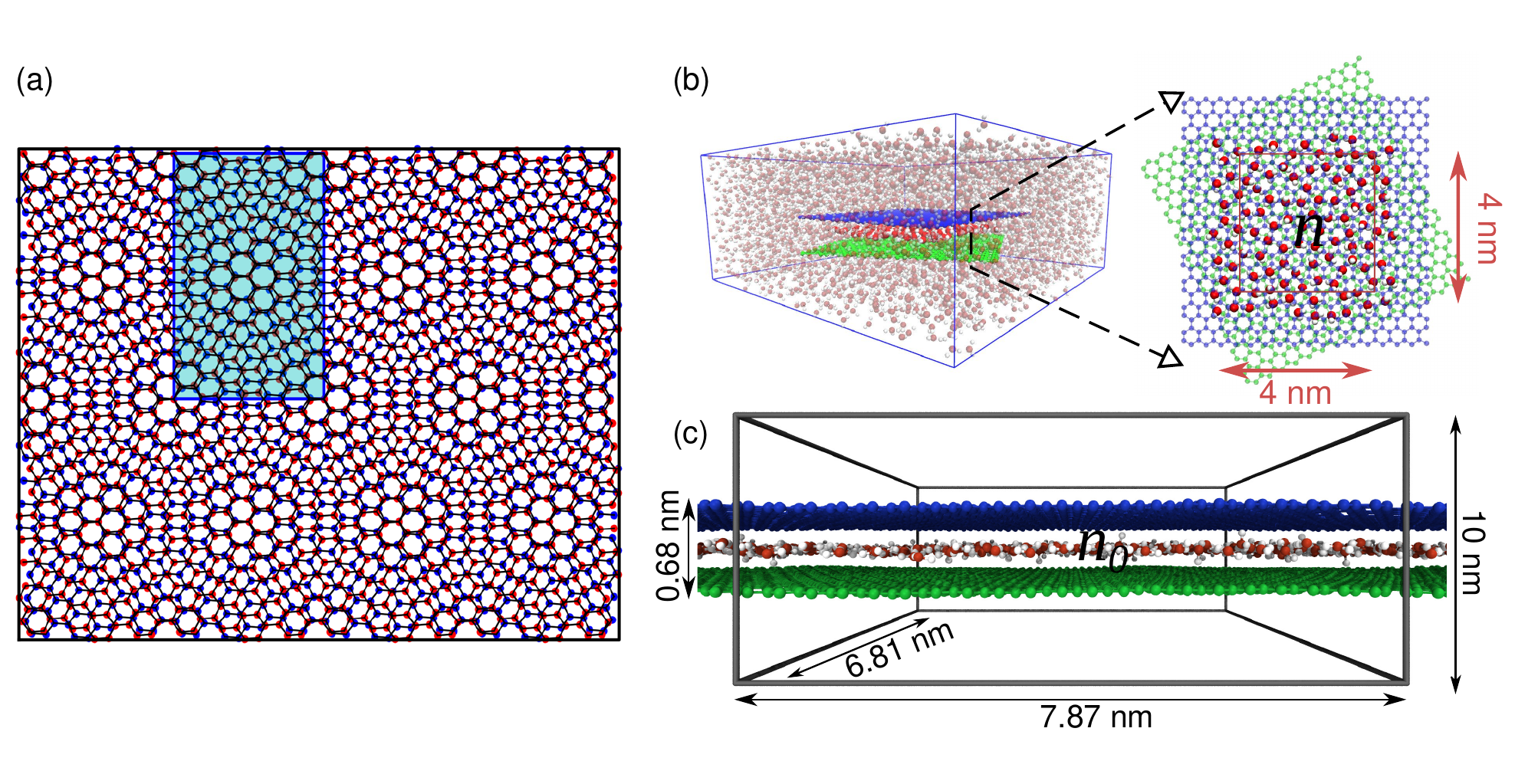}
    \caption{
    (a) The blue rectangular patch in the left image shows a single Moir\'{e} unit and the entire region is a schematic representation for $\SI{9.4}{\degree}$ tBLG. 
    (b) On the top right, a snapshot of finite rigid tBLG immersed in a bath of water rotated at a specific relative twist angle ($\theta$) is given, and (c) on the bottom right, a snapshot of the total system used as an initial structure of periodic confined water-graphene system is given from side view. The water filled periodic tBLG sheets at densities $(n_0)$ are initially build with $\SI{0.68}{\nano\meter}$ separation between the two graphene sheets, and a total box size of 10 nm was used to prevent any periodic interaction with the water layers from the image cells. 
    }
    \label{fig:schematic}
\end{figure}

We perform atomistic MD simulations of monolayer water confined inside the slit-pore tBLG sheets rotated at different twists angles as shown in Fig.~\ref{fig:schematic}.
To generate the infinite graphene sheets for periodic tBLG system, we note that graphene exhibits a two dimensional honeycomb-like hexagonal lattice structure with lattice constant $a = \SI{2.46}{\angstrom}$.
Adding an identical, but rotated graphene layer on top of an existing graphene sheet produces large scale crystallographic structures called Moir\'{e} structures as shown in Fig.~\ref{fig:schematic} a.
The smallest size of a supercell that can contain at least one Moir\'{e} is inversely proportional to twist angle.
Therefore small angle tBLG systems become prohibitively large for simulation.
This restricts simulations with infinitely periodic tBLG to only some twist angles where the number of carbon atoms ($C$), forming a supercell, is relatively small.
We focus our studies on such commensurate twist angles defined by,
\begin{equation}
    \cos\theta(m,r) = \frac{3m^2+3mr+r^2/2}{3m^2+3mr+r^2}
\end{equation}
where, $m$ and $r$ are co-prime positive integers\cite{struc_moire}.
The lattice structures of commensurate systems are also hexagonal.
Their primitive lattice translation vectors are given by,
\begin{gather}
    \vec{c_1} = m \vec{a_1} + (m+r) \vec{a_2} \\
    \vec{c_2} = -(m+r) \vec{a_1} + (2m+r) \vec{a_2}
\end{gather}
where, $\vec{a_1}$ (=$\sqrt{3}a_{CC}(\frac{\sqrt{3}}{2}, \frac{1}{2})$) and $\vec{a_2}$ (=$\sqrt{3}a_{CC}(\frac{\sqrt{3}}{2}, -\frac{1}{2})$) are the basis vectors for monolayer graphene ($a_{CC}$ being the carbon-carbon bond length of graphene) and $r$ obeys the relation, $\text{gcd}(r, 3) = 1$.
The number of basis atoms in a primitive unit cell can also be specified in terms of $m$ and $r$\cite{twbl-gr_struct_n}\cite{commensurate_gr}.
Triclinic supercells of tBLG comprising $4$ Moir\'{e} structures ($6$ in cases of $\theta = \SI{21.8}{\degree}, \SI{27.8}{\degree}$) are generated for these commensurate angles inside a simulation box that preserves their periodicity. 
The two layers that are initially kept at a separation of $\SI{6.8}{\angstrom}$ (almost twice their equilibrium interlayer separation $H = \SI{3.412}{\angstrom}$)\cite{gargiulo2017structural}.
At this separation, the nanoslit pore allows only a monolayer of water inside it \cite{raghav2015molecular}.

We executed multiple simulations using SPC/E and TIP4P/2005 water model \cite{berendsen1984molecular, abascal2005general} and finite tBLG sheets at chosen commensurate twist angles: $\SI{0.0}{\degree}$, $\SI{3.5}{\degree}$, $\SI{5.1}{\degree}$, $\SI{9.4}{\degree}$, $\SI{16.4}{\degree}$, $\SI{21.8}{\degree}$, $\SI{26.0}{\degree}$, $\SI{27.8}{\degree}$, $\SI{29.4}{\degree}$. 
The carbon $(C)$ atoms of tBLG are modelled as uncharged Lennard-Jones (LJ) particles using AMBER FF10 force fields of type \textit{CA} \cite{mukherjee2007structure, mukherjee2007strong, homeyer2006amber}.
These atoms are held rigidly to their initial position using a harmonic restraining force (force constant $\SI{100}{kcal \per\mole\per\angstrom\squared}$).
SHAKE algorithm \cite{ryckaert1977numerical} was used to constrain all the bonds and angles involving hydrogen of every water molecule.
The initial system was minimised using steepest descent followed by conjugate gradient algorithms for $5000$ steps.
Then the system was heated to $\SI{300}{K}$ using Langevin thermostat in steps of $\SI{30}{K}$ \cite{schneider1978molecular}. 
Once the desired temperature is reached, we perform \textit{NPT} equilibration at $\SI{300}{\kelvin}$ and at $\SI{1}{atm}$ pressure using Nosé-Hoover thermostat and barostat respectively \cite{nose1984unified, hoover1985canonical, martyna1994constant}.
The value of temperature coupling constant used is $\SI{200}{\femto\second}$ and that of pressure coupling constant is $\SI{2}{\pico\second}$.
In this ensemble, the system gradually equilibrates and the density effectively corresponds to $\SI{1}{atm}$ pressure.
For SPC/E water model,
a cut-off of $\SI{9}{\angstrom}$ is used for short-range interactions, beyond which, a switching function is used to decay the potential to zero at a distance of $\SI{10}{\angstrom}$.
For TIP4P/2005 water model, a global cut-off of $\SI{10}{\angstrom}$ is used.
Long-range electrostatic energies are calculated in the reciprocal $k-$space using PPPM solver \cite{hockney1988computer} with an accuracy of $10^{-4}$. 
The system density equilibriates after $\sim \SI{500}{\pico\second}$ in the total $\SI{1}{\nano\second}$ NPT run. 
The density equilibrated system was then subjected to a \textit{NVT} simulation at $\SI{300}{\kelvin}$ for another $\SI{20}{\nano\second}$.
All the simulation protocols are performed using LAMMPS \cite{plimpton1995fast} with velocity-verlet integrator using a  time step of  $\SI{2}{\femto\second}$.
This entire process allows sufficiently large number of exchanges of water molecules between the water bath (reservoir) and the nanochannel, establishing a chemical equilibrium among the two regions \cite{mondal2021anomalous}.
The schematic for these simulations is shown in Fig.~\ref{fig:schematic}b.
After this \textit{NVT} simulation, the number of confined water molecules within $\SI{4}{\nano\meter} \times \SI{4}{\nano\meter}$ from the centre of the tBLG immersed in water bath becomes constant.
The density ($n$) (confined water molecules per unit area) obtained for each twist angle in the above described finite sheet simulation was used as target filling density for the infinite periodic sheets where the two twisted graphene sheets are now periodically connected to their images in neighbouring simulation cells in x and y direction (shown in Fig.~\ref{fig:schematic}c).

Using the calculated filling densities as described above at different twist angles, water molecules are randomly inserted inside the slit-pore of infinite tBLG sheets at desired angles.
To remove overlaps of atomic coordinates and bad contacts, molecules lying within \SI{1}{\angstrom} radius are deleted and subsequently, further insertion attempts are made until the desired number density is achieved inside the channel.
This step is performed using available LAMMPS commands (\textit{region, molecule, create\_atoms, delete\_atoms, reset\_mol\_ids}) and automated bash scripting.
The final confined system generated using different choices of filling densities are energy minimized and simulated for $\SI{20}{\nano\second}$ in NVT ensemble using the same timestep and integrator, as used in the case of finite sheets described above.
Here, the carbon atoms of the infinite tBLG sheets are no longer restrained to their initial positions during the simulation. 
They are therefore free to relax, maintaining the infinite periodicity with the adjacent cells.

\section{III. Results and Discussion}
\label{section_results_discussion}
\subsection{A. Confinement Area, Height and Water Density}
\label{confined_water_density}
\begin{figure}
     \centering
     \begin{subfigure}[b]{0.4\textwidth}
         \centering
         \includegraphics[width=\textwidth]{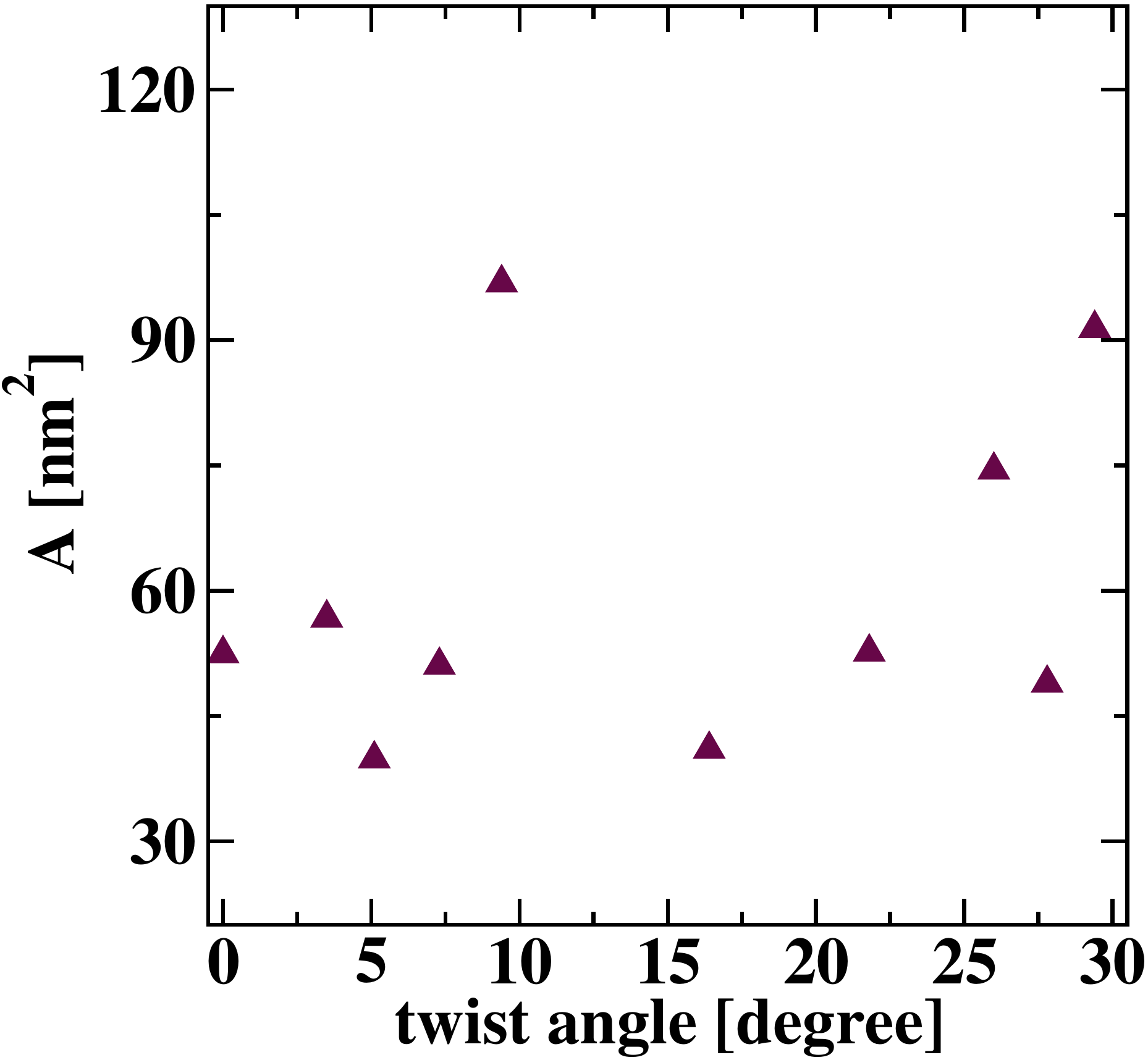}
         \caption{}
         \label{}
     \end{subfigure}
     %\hfill
     \begin{subfigure}[b]{0.4\textwidth}
         \centering
         \includegraphics[width=\textwidth]{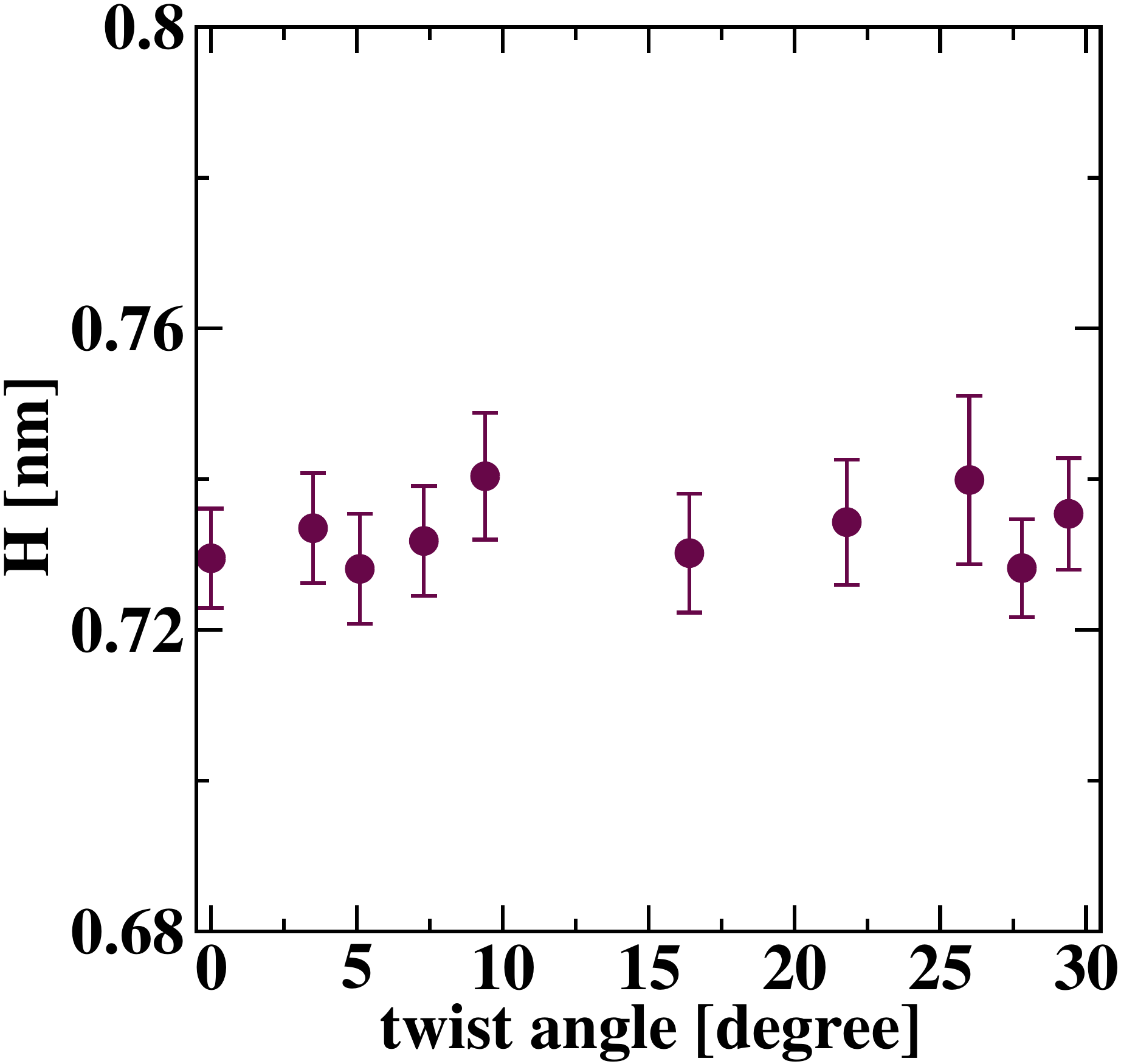}
         \caption{}
         \label{}
     \end{subfigure}
     \caption{On the left (a), areas (A) used for each twist of periodic tBLG sheets is shown. 
     Areas are chosen such that at least 4 Moir\'{e} units are there for every twist.
     On the right (b), the average distance (H) between the graphene sheets is plotted as a function of twist. 
     We see that H remains almost constant over all the twist angles explored here.}
     \label{fig:area_height}
\end{figure}

The confined area used for each twist varies due to size differences in Moir\'{e} unit, and the chosen area which is an integer multiples of this Moir\'{e} unit is shown in Fig.~\ref{fig:area_height}a for every twist of tBLG.
As the restraint forces of the graphene atoms of these periodic tBLG are released, they are therefore able to relax along the sheet's normal direction ($z$ direction) with the water layer in between.
For the case of confined SPC/E water, we plot this distance between the two graphene sheets of tBLG in Fig.~\ref{fig:area_height}b. 
For all the twist angles the height can be seen to be around at $\sim 7.3$ \AA\ with a negligible variation ($\sim 0.125$ \AA). 

\begin{figure}[t!]
    \centering
    \includegraphics[width=0.8\textwidth]{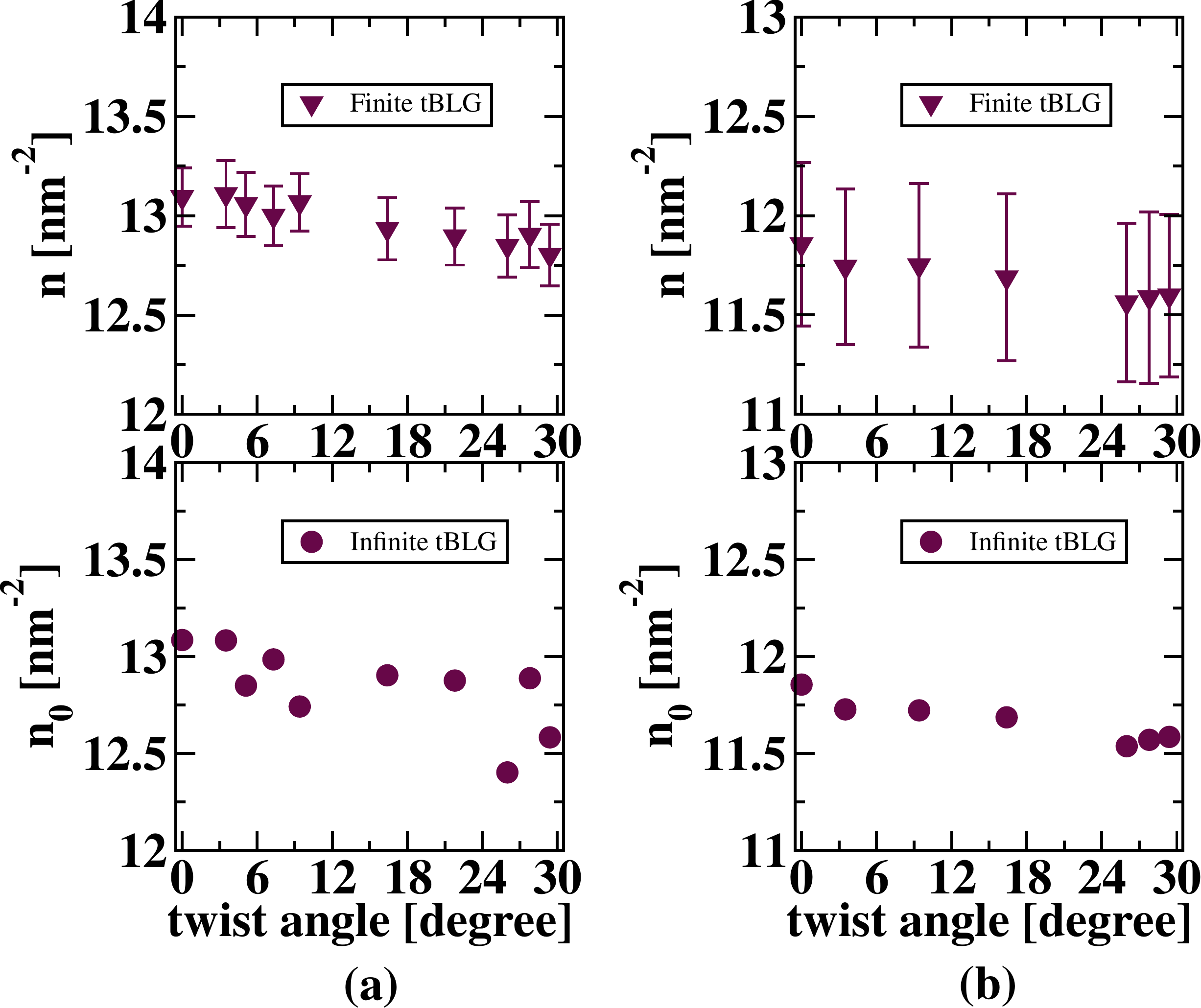}
    \caption{
    Top row plots shows the obtained water density from finite tBLG simulation in water bath. The plots at the bottom rows show the final densities used to build the confined water system by inserting water molecules within the periodic tBLG sheets discretely for (a) SPC/E, (b) TIP4P/2005 water model.
    The densities are calculated by counting the number of water molecules sandwiched within $4 \times \SI{4}{\per\nano\meter\squared}$ region of the tBLG and averaged over the production runs. 
    }
    \label{fig:confined_water_density}
\end{figure}

The target water density inside the channel, obtained from finite sheet simulations ($n$) at different twist angles are shown in Fig.~\ref{fig:confined_water_density} (finite tBLG), while the exact achieved water densities $(n_0)$ inside periodic tBLG, keeping $n$ as target, is given in the bottom of the same figure (infinite tBLG).
The density plot shows that the obtained density ($\sim 12.7-\SI{13.1}{\per\nano\meter\squared}$ for SPC/E and $\sim 11.5-\SI{12}{\per\nano\meter\squared}$ for TIP4P/2005) has negligible difference with twist angle. 
For our filling densities chosen from finite tBLG simulations, the water remain as a monolayer.
Hence, it can be said that the interlayer force of graphene atoms are pacified with a single monolayer of water between them so that the effect of different graphene stacking due to twist is subdued. 
With our knowledge on final confinement dimension and density of water that resides within, we seek to look for the presence of any structural changes that might occur in the confined water molecules.

\subsection{B. Confined water structure}
\label{confined_water_structure}
\begin{figure}[ht]
    \centering
    \includegraphics[width=0.5\textwidth]{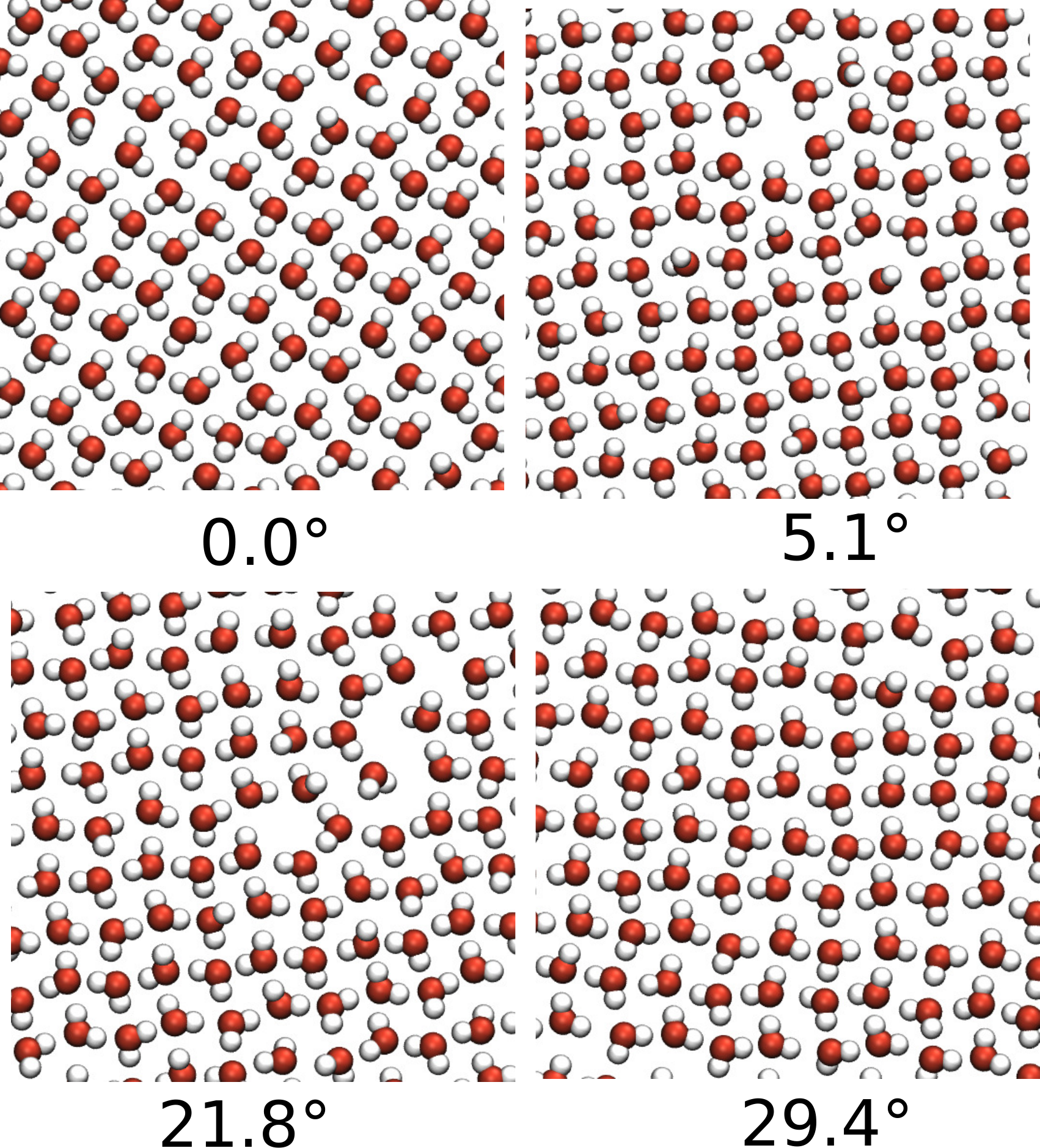}
    \caption{
    Snapshots of a region of confined SPC/E water inside tBLGs of four different angles out of all the twist angles studied in this work. 
    The water structures showing square ice ordering inside the channel for all the four twist angles shown.
    }
    \label{fig:confined_water_snapshot}
\end{figure}

For our obtained densities in the case of SPC/E monolayer water, square-ice ordering was observed at all twists and snapshots of four such ordered water structure is shown in Fig.~\ref{fig:confined_water_snapshot}. 
This corroborates with the experimental observation of square ice ordering of confined water in graphene. 
However, for TIP4P/2005, no such square-ordered structure was obtained at any twist angle. 
The inability of this model to show square ordering was already known for \SI{0}{\degree} un-twisted graphene confinement \cite{majumdar2021dielectric}, and now it can be generalized to other twist angles as well.

\begin{figure}[ht]
    \centering
    \includegraphics[width=\textwidth]{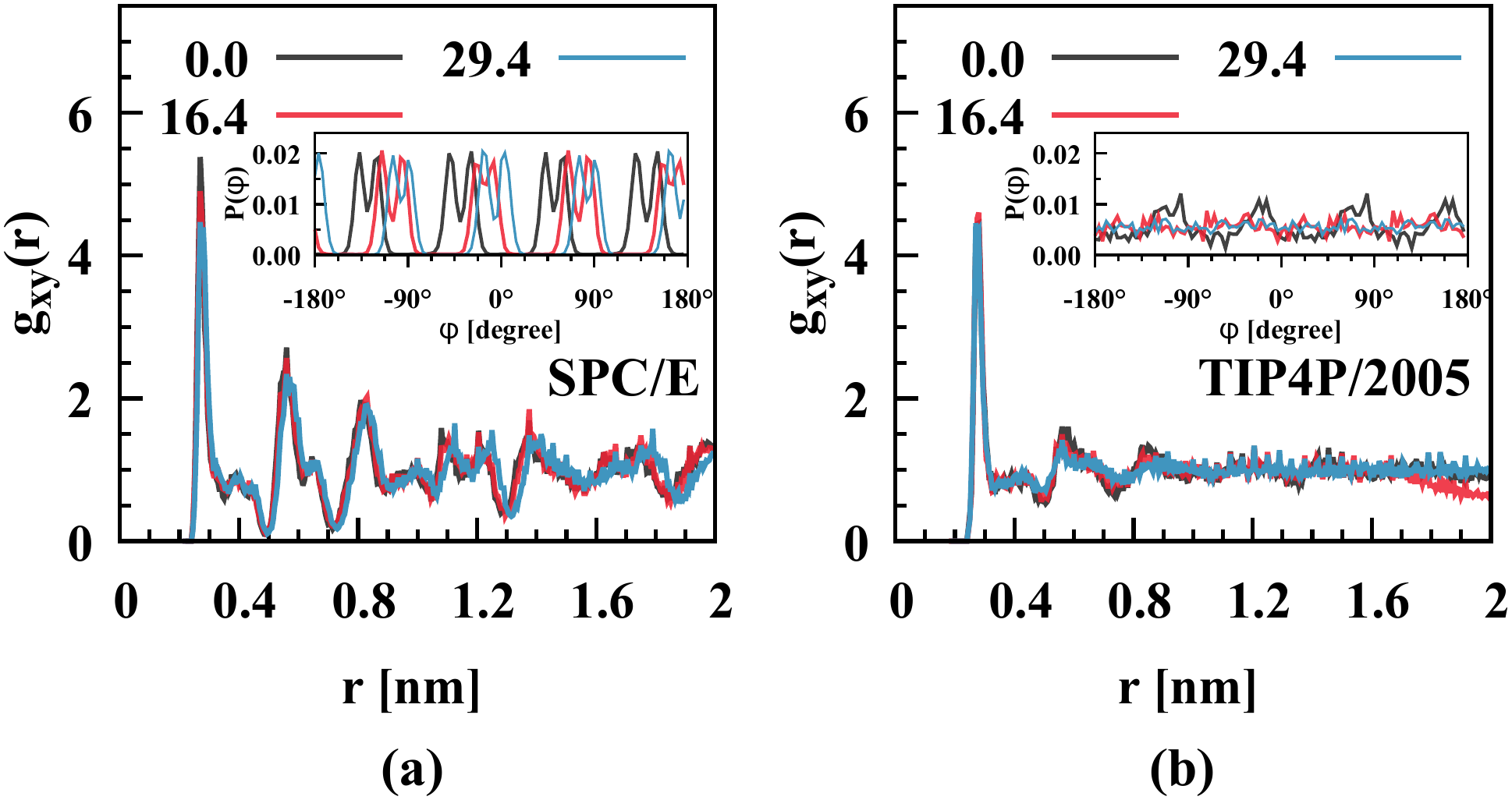}
    \caption{
    The radial distribution of the in-plane confined oxygen atoms are given for three twist angles for (a) SPC/E and (b) TIP4P/2005 water models}. They all show the same radial distribution. In inset, the angle distribution of $\phi$ used in Eq. \ref{equation:order_parameter} is given.
    \label{fig:twist_rdf_split.pdf}
\end{figure}

%with a square lattice parameter of 2.8 \AA. % HEY
$\Psi_4$ order parameter is used to quantify this square ordering. 
It is defined as,
\begin{equation}
     \Psi_{4} = \left< \left| \frac{1}{N} \sum \limits_{m=1}^{n} \frac{1}{n_{b}} \sum \limits_{n=1}^{n_{b}} e^{i k \phi_{mn}} \right| \right>\\
     \label{equation:order_parameter}
\end{equation}
where, $n$ runs over the nearest neighbours of $m^{th}$ oxygen atom,\
$N$ is the total number of confined oxygen atom, \
$n_{b}$ is the total number of nearest neighbours identified by a distance cut-off,\
and $\phi_{mn}$ is the angle between the line joining two oxygen atoms of neighbouring molecules with respect to a planar axis.
The cut-off for nearest neighbours is taken to be the first minima of planar $O-O$ radial distribution function, $g_{xy}(r)$, of confined water molecules.
This radial distribution function (RDF) which is plotted in Fig.~\ref{fig:twist_rdf_split.pdf} is computed using,
\begin{equation}
    g_{xy}(r_{i}) = \frac{1}{n_0 N} \frac{N_{i}}{2\pi r \Delta r}
\end{equation}
where, $N_i$ is the number of particles within an annular ring of radius $r_i$ and thickness $\Delta r$, $n_0$ is the number density and $N$ is the total number of  water molecules \cite{cuadros1992radial}.
\begin{figure}[ht]
    \centering
    \includegraphics[width=0.9\textwidth]{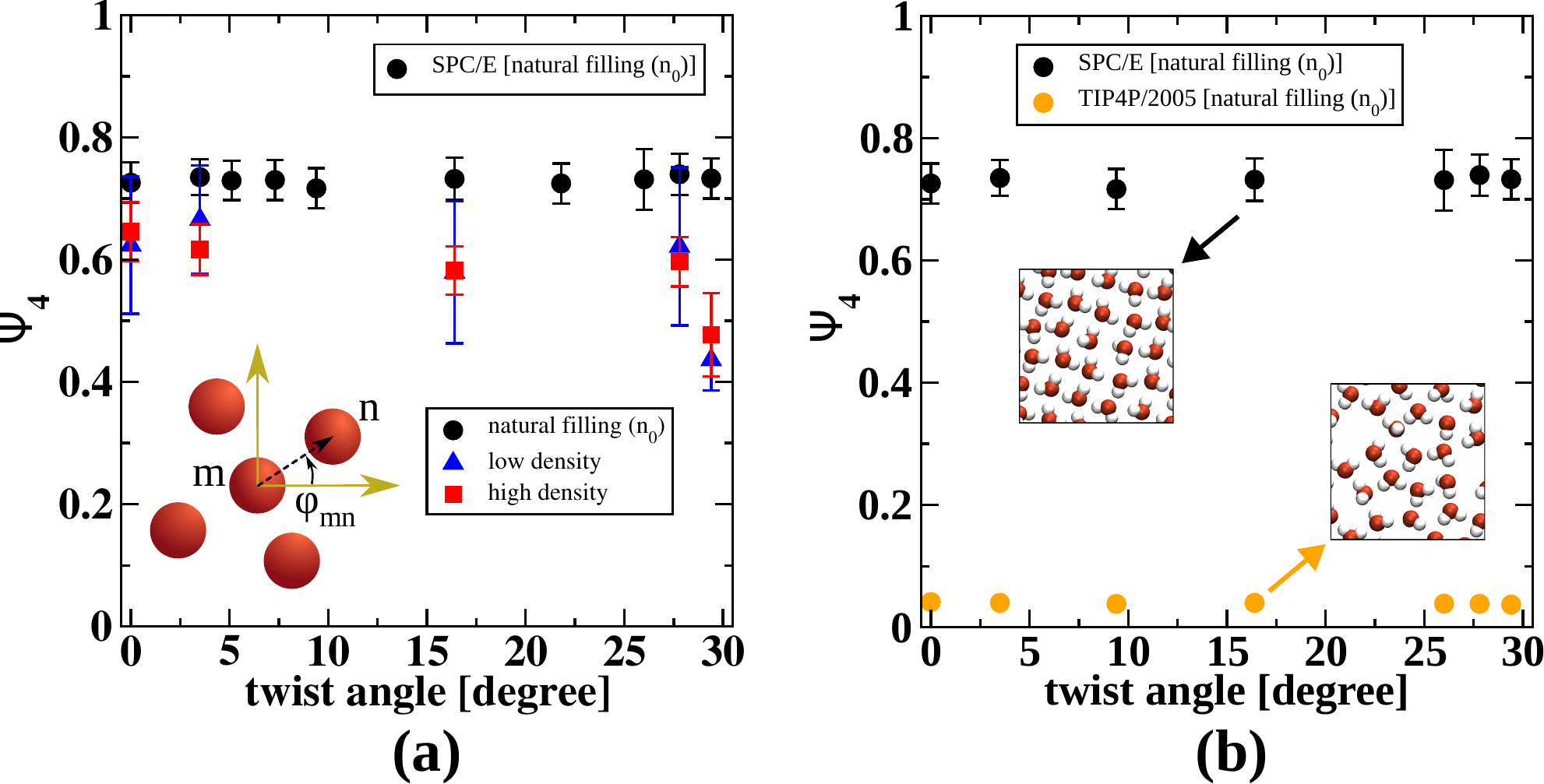}
    \caption{
    $\Psi_4$ order parameter (Eq. \ref{equation:order_parameter}) as a function of twist angle for 3 different filling densities  in the case of (a) SPC/E water model.
    $n_0$ gives higher ordering than a low or high density filling $n_V$ and $n_D$ respectively. 
    A schematic diagram shows the definition of angle $\phi_{mn}$ between two atoms $m$ and $n$,  used in computing $\Psi_4$. 
    (b) Comparison of $\Psi_{4}$ obtained using SPC/E and TIP4P/2005 for natural filling density with snapshots corresponding to $\theta = \SI{16.4}{\degree}$.
    }
    \label{fig:twist_ktic_50ns_inset.pdf}
\end{figure}
$\Psi_4$ versus twist angle is shown in Fig.~\ref{fig:twist_ktic_50ns_inset.pdf}. 
The high square-ice ordering of the confined water molecules is evident from the high value of $\Psi_4$, which is also very much invariant of the twist angles.
This shows that the square-ordering of tBLG confined monolayer water is agnostic of the relative rotational orientation of the two confining planes. 
This invariance can be seen for TIP4P/2005 model as well.
From Fig.~\ref{fig:twist_ktic_50ns_inset.pdf}(b) it can be seen that in comparison to SPC/E, the values of $\Psi_4$ in the case of this 4-point water model is very low.

Distribution of $\phi$ used in Eq. \ref{equation:order_parameter} is given as an inset of Fig.~\ref{fig:twist_rdf_split.pdf} for three twist angles, and it shows an imperfect square ordering. 
This is because, a perfect square ordering would have corresponded to peaks separated by $\SI{90}{\degree}$ without any peak split, while we have peak splits present in the distribution.
In a previous work we had obtained such peak splits in the distribution of SPC/E monolayer water, where we connected the distribution with a mixed combination of purely rhombic and square ordering, and justified with what had been previously found in experiment \cite{majumdar2021dielectric, zhou2015observation, wang2015wang}. 
The relative phase shift in the distribution of $\phi$ happens because the complete water layer can have a random overall rotational shift with respect to the planar axis used to calculate $\Psi_4$ (here, $x$-axis).

\subsection{C. Confined water static permittivity}
\label{confined_water_dielectric} 
Structural analysis in the previous section highlighted the organization of confined water molecules as a collection.
Although internal structural variation was not observed for different twist angles, each water dipole can still fluctuate inside the slit-pore in collaboration with its neighbour.
Suppose the effect of twist on the water-graphene interaction at room temperature is significant. In that case, the fluctuation of water dipoles is expected to bear its signature since the average fluctuation of water dipole is directly proportional to static permittivity \cite{varghese2019effect}.
This static permittivity of the confined water has been addressed in many recent works which showed increasing anomalous nature of water as it is suppressed from bulk to confinement \cite{senapati2001dielectric,fumagalli2018anomalously, bonthuis2012profile, schlaich2016water, ruiz2020quantifying, jalali2020out}.
The perpendicular permittivity of nanoconfined water ($\epsilon_{\perp}$) is a response function which is calculated from the fluctuation of the dipole moment along the direction of confinement. 
However, in the case of extreme confinement within slab geometry, the breaking of isotropy in the $z$ direction makes the dielectric function depended on $z$; denoted by $\epsilon_{\perp}^{-1}(z)$. 
In this work, the effective dielectric constant is computed in two ways: one which uses the integral of $\epsilon_{\perp}^{-1}$ (from the work of Stern and Feller, represented as SF) \cite{stern2003calculation} and another, which uses the analytical expression of this computed integral (represented as BM) (similar to what is given by Mondal \emph{et. al.} \cite{mondal2021anomalous} for simulations done with pseudo-2D particle mesh Ewald summation).

In method SF, the inhomogeneous $\epsilon_{\perp}^{-1}$, as a function of $z$, is computed from the equation \cite{stern2003calculation, bonthuis2012profile},
\begin{equation}
    {(\epsilon^{SF}_{\perp}(z))}^{-1} = 1 - \frac{c_{\perp}(z)}{\epsilon_{0} k_{B}T + \frac{C_{\perp}}{V}}
    \label{eq:permittivity_hansen_main}
\end{equation}
\begin{equation}
    \epsilon^{SF}_{\perp} = \frac{(L_2 - L_1)}{\int^{L_2}_{L_1} {(\epsilon^{SF}_{\perp}(z))}^{-1} dz}
    \label{eq:permittivity_hansen}
\end{equation}
where, $c_{\perp}(z)=\langle m_{\perp}(z) m_{\perp} \rangle_{0} -  \langle m_{\perp}(z)\rangle_{0}\langle m_{\perp} \rangle_{0}$ and $C_{\perp} = A\int c_{\perp}(z)dz$.
The brackets $\langle$ $\rangle_0$ denote ensemble averages without an externally applied electric field. 
$m_{\perp}(z)$ denotes the perpendicular component of water dipole density at height $z$ from the bottom sheet, and $m_{\perp}$ is the integration of $m_{\perp}(z)$ over the slit width.
$L_1$ and $L_2$ are the positions of the bottom sheet and top sheet, respectively. 
In the absence of any external electric fields, the fluctuations of $m_{\perp}$ give rise to a net polarization $(M_{\perp})$, which can be directly used to compute the effective $\epsilon_{\perp}$ as in method BM.

\begin{equation}
    (\epsilon^{BM}_{\perp})^{-1} = \frac{A}{V} \int^{L_2}_{L_1} {(\epsilon^{SF}_{\perp}(z))}^{-1} dz 
    = \frac{1}{V}\int^{L_2}_{L_1} A dz -  \frac{1}{\epsilon_{0} k_{B}T + \frac{C_{\perp}}{V}} \int^{L_2}_{L_1} A c_{\perp}(z) dz \\
    \label{eq:permittivity_biman_mondal_main}
\end{equation}
\begin{equation}
    \epsilon^{BM}_{\perp} = \frac{1}{(\epsilon^{BM}_{\perp})^{-1}} = 
     \left[1 - \left(\frac{1}{V}\right)\frac{C_{\perp}}{\epsilon_{0} k_{B}T + \frac{C_{\perp}}{V}} \right]^{-1}
    \label{eq:permittivity_biman_mondal}
\end{equation}

where, $\beta = 1/(k_{B} T)$, $T$ is the temperature of the system.
The confined volume of the nanochannel is given by, $V = A(H - \sigma_{CO})$, where $A$ is the area of tBLG, $H$ is the interlayer spacing and $\sigma_{CO}$ is the cross-interaction LJ parameter between Carbon and Oxygen atoms calculated using Lorentz-Berthelot combination rule.
\begin{figure}[ht]
    \centering
    \includegraphics[width=0.9\textwidth]{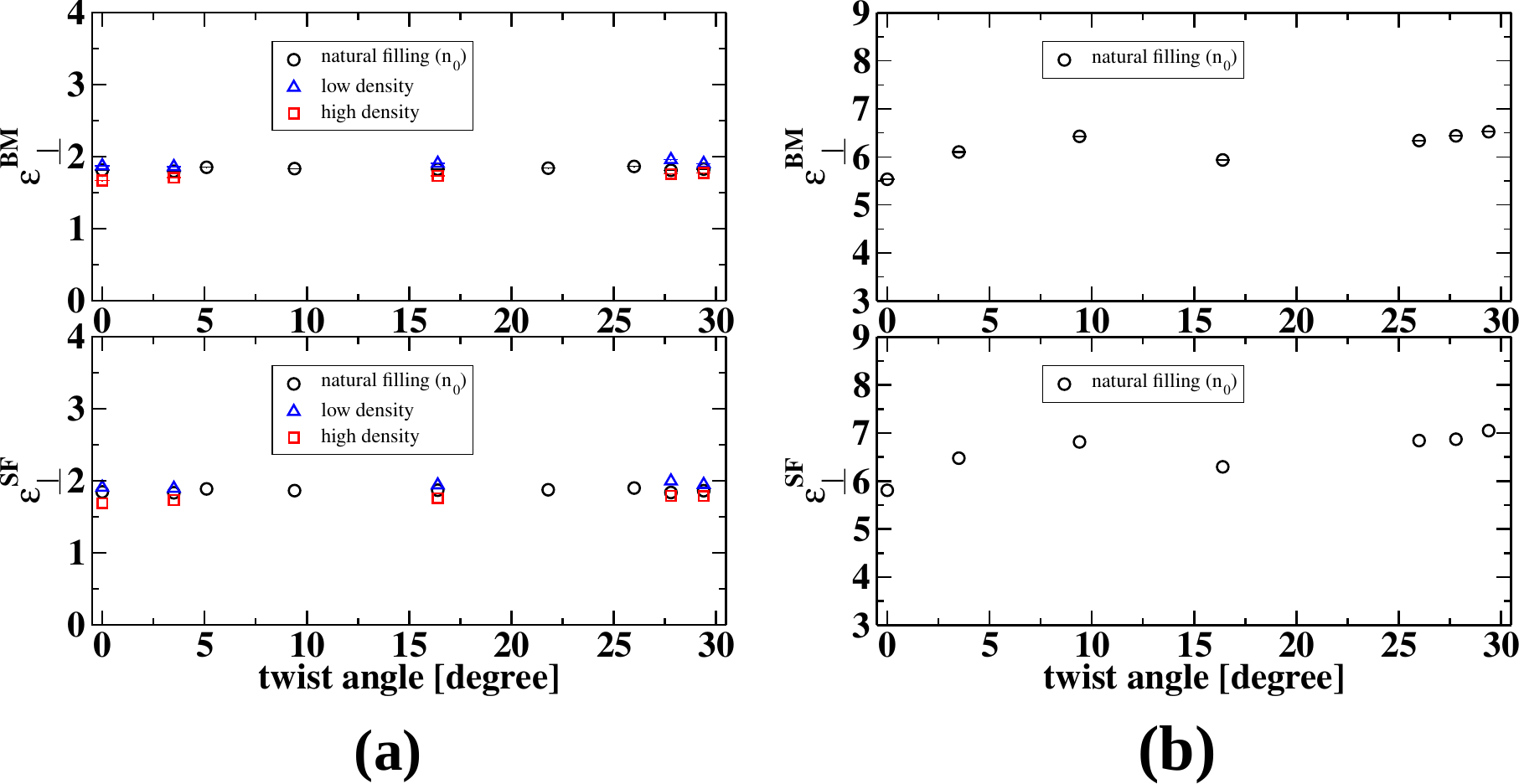}
    \caption{
    The perpendicular static permittivity of confined water is shown for $n_0$, low density and high density.
    It can be seen that the dielectric constant remains reasonably unaffected with the twists of the tBLG.
    (a) shows the values obtained for SPC/E model. It also comprises the case of low density and high density filling. 
    (b) shows the values obtained for TIP4P/2005 model. 
    The disordered dipoles of TIP4P/2005 water give rise to a higher dielectric constant value than SPC/E water.
    }
    \label{fig:twist_epsilonAvgPerp_50ns.pdf}
\end{figure}
The values obtained using the above equations are shown in Fig.~\ref{fig:twist_epsilonAvgPerp_50ns.pdf}. 
The top row plots show the permittivity values obtained from Eq. \ref{eq:permittivity_biman_mondal} $(\epsilon^{BM}_{\perp})$, while the bottom row plots show the values that are obtained by numerical integration of Eq. \ref{eq:permittivity_hansen_main} (SF method) as given in Eq. \ref{eq:permittivity_hansen}  $(\epsilon^{SF}_{\perp})$.
From the plots, it is evident that the dielectric constant of the confined water do not change with twist angle. 
For the respective models, the values are constant.
Owing to lack of square-ordering, TIP4P/2005 water can have more fluctuations inside the channel. 
This results in a higher dielectric constant compared to monolayer SPC/E water. 
Thus, the static permittivity of confined water in tBLG do not bear any significant signature of twist angle of tBLG.
However, one possibility remains, where maintaining similar dipole fluctuation and structural ordering, the confined water molecules may flow distinctly with varying twist angles.

\subsection{D. Friction coefficient and slip length of confined water }
\begin{figure}[t!]
\centering
\includegraphics[width=\textwidth]{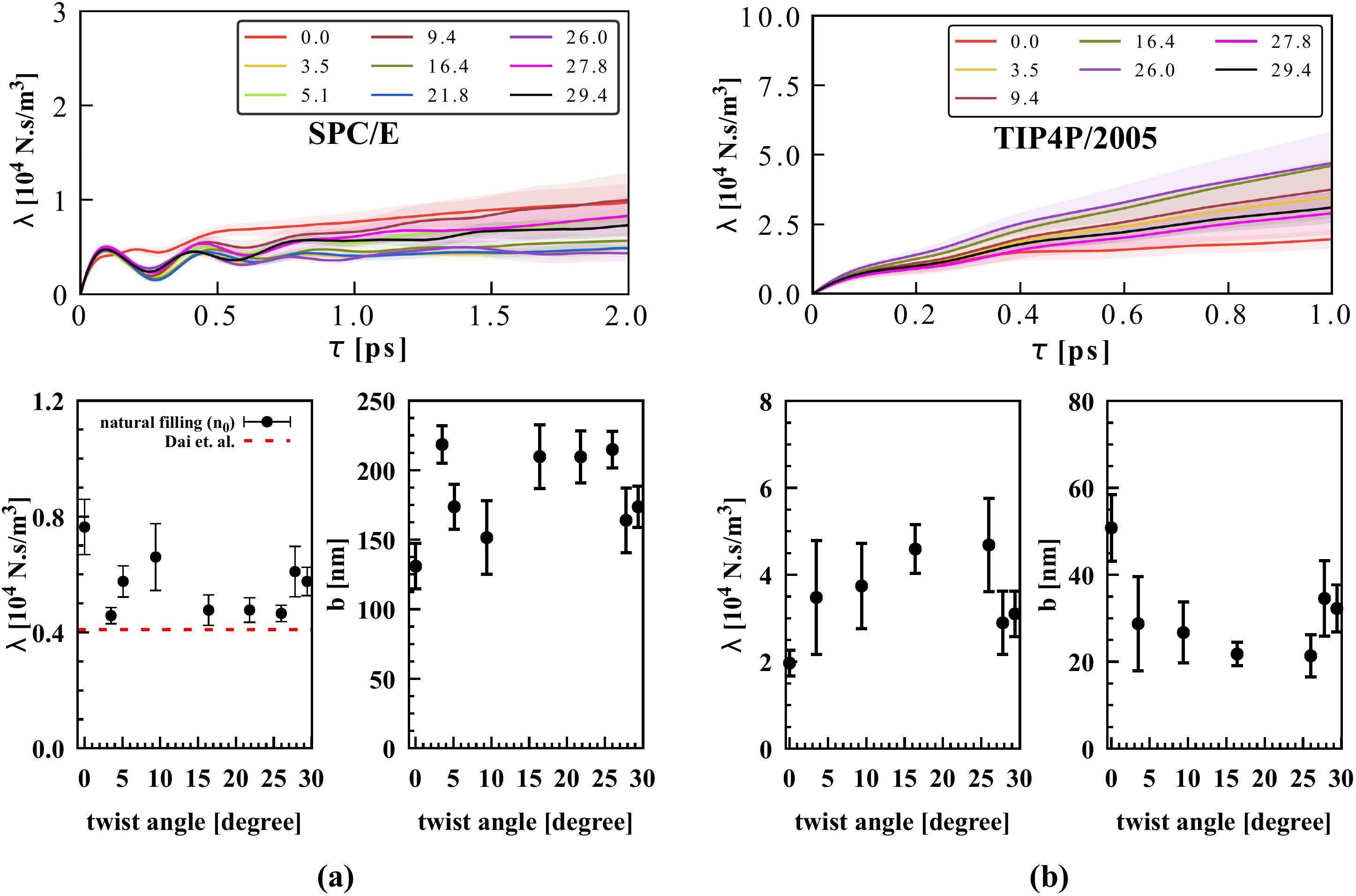}
\caption{
    Friction coefficient $(\lambda)$ and slip length $(b)$ of (a) SPC/E and (b) TIP4P/2005 waters inside restrained tBLG sheets
    The top row shows the variation of $\lambda$ with time and the bottom row shows the final values.
    For SPC/E water model we compare our obtained values with previously reported values from the work of Dai. et. al. using SPC/E water. \cite{dai2016interfacial}
    }
\label{fig:friction_comparison}
\end{figure}

The friction coefficient of confined water $(\lambda)$ is calculated from the auto-correlation of force experienced by the water molecules from the graphene sheets as given by the following equation,
\begin{equation}
    \lambda = \frac{1}{2 A k_B T} \int_0^{\infty} \left < F(0) F(t) \right > dt
    \label{eq:friction}
\end{equation}
where $F$ is the force between water and restrained graphene atoms, $A$ and $T$ are temperature of simulation and area of confinement respectively, and $k_B$ is the Boltzmann constant.
This specific calculation was done in NVE ensemble for $\SI{100}{\pico\second}$ and forces were dumped at $\SI{1}{\femto\second}$ interval.
Five independent runs were performed for each twist angle, whose data was then used for computing $\lambda$ for different twist angles shown in Fig.~\ref{fig:friction_comparison}.
The final value of $\lambda$, we compute the slip length using the relation $b = \eta / \lambda$ \cite{bocquet2007flow, tocci2014friction} shown in the bottom row of Fig.~\ref{fig:friction_comparison}. 
To our understanding, the random kicks experienced by confined water molecules leads to slow convergence of $\lambda$. 
The slow convergence can also be attributed to high standard deviation in potential energy of the different AA and AB stacking of tBLG, as confirmed in earlier work of Wagemann \emph{et. al.} \cite{wagemann2020quantifying}. 
Our calculated friction coefficient also agrees well with  Dai et. al. \cite{dai2016interfacial}.
The SPC/E water model shows a significant dependence of $\lambda$ for small twist angles while the TIP4P/2005 model shows little variation.
This qualitative difference between the two water models arises from the difference in their ordering.
Consequently, the slip-length which has inverse relation to friction also shows a qualitative difference for the two water models.
Despite the mismatch, it is evident from the time series of $\lambda$ that the time taken for convergence is greatly dependent on the twist angle.

\subsection{E. A case of high and low density filling in the nano-channel}

As described in the method section, till now, we discussed the results based on the water density that was obtained from finite tBLG (restrained) simulation in a water bath. 
Since the average filling density came out to be same for a particular water model, the calculations discussed earlier should capture differences that can arise only from twists.
However, the results above show that twist induces almost negligible effect on confined water properties.
Therefore what remains is to calculate the same quantities when the density is manually chosen slightly lower and higher than the obtained values from finite tBLG in water bath simulation.

\begin{figure}[ht]
    \centering
    \includegraphics[width=0.4\textwidth]{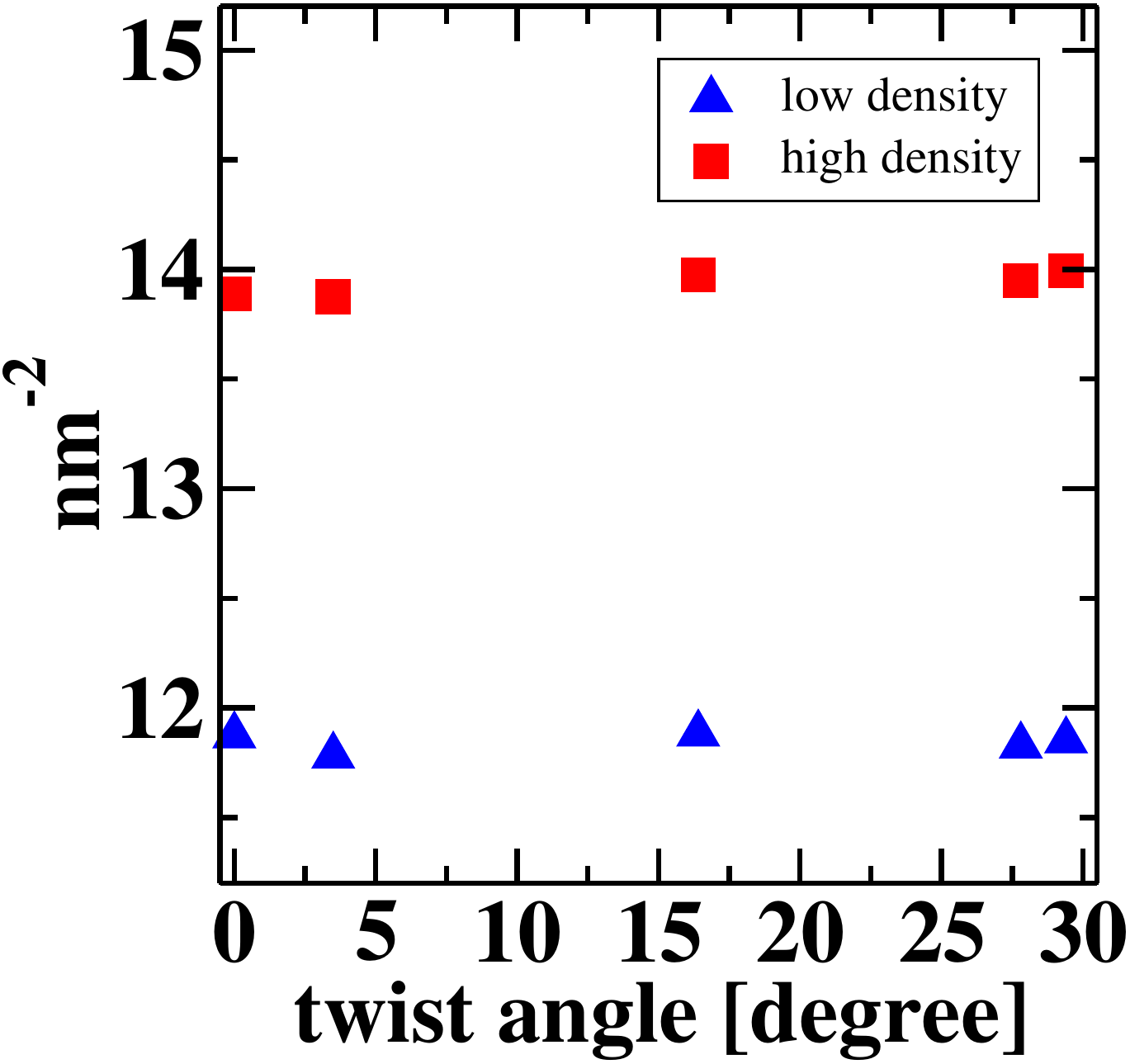}
    \caption{
    The exact density values used to fill the confined region of tBLG in case of high density filling $n_D$ and low density filling $n_V$ is plotted for the five twist angles explored in this section.
    }
    \label{fig:high_low_density.pdf}
\end{figure}

We explored two densities for SPC/E and investigated properties at five twist angles ($\SI{0}{\degree}$, $\SI{3.5}{\degree}$, $\SI{16.4}{\degree}$, $\SI{27.8}{\degree}$ and $\SI{29.4}{\degree}$): a high density  ($\sim \SI{14}{\nano\meter}^{-2}$) and a low density ($\sim \SI{12}{\nano\meter}^{-2}$), inside infinite periodic tBLG. 
The exact densities used for SPC/E water model are plotted in Fig.~\ref{fig:high_low_density.pdf}.

\begin{figure}[ht]
    \centering
    \includegraphics[width=0.8\textwidth]{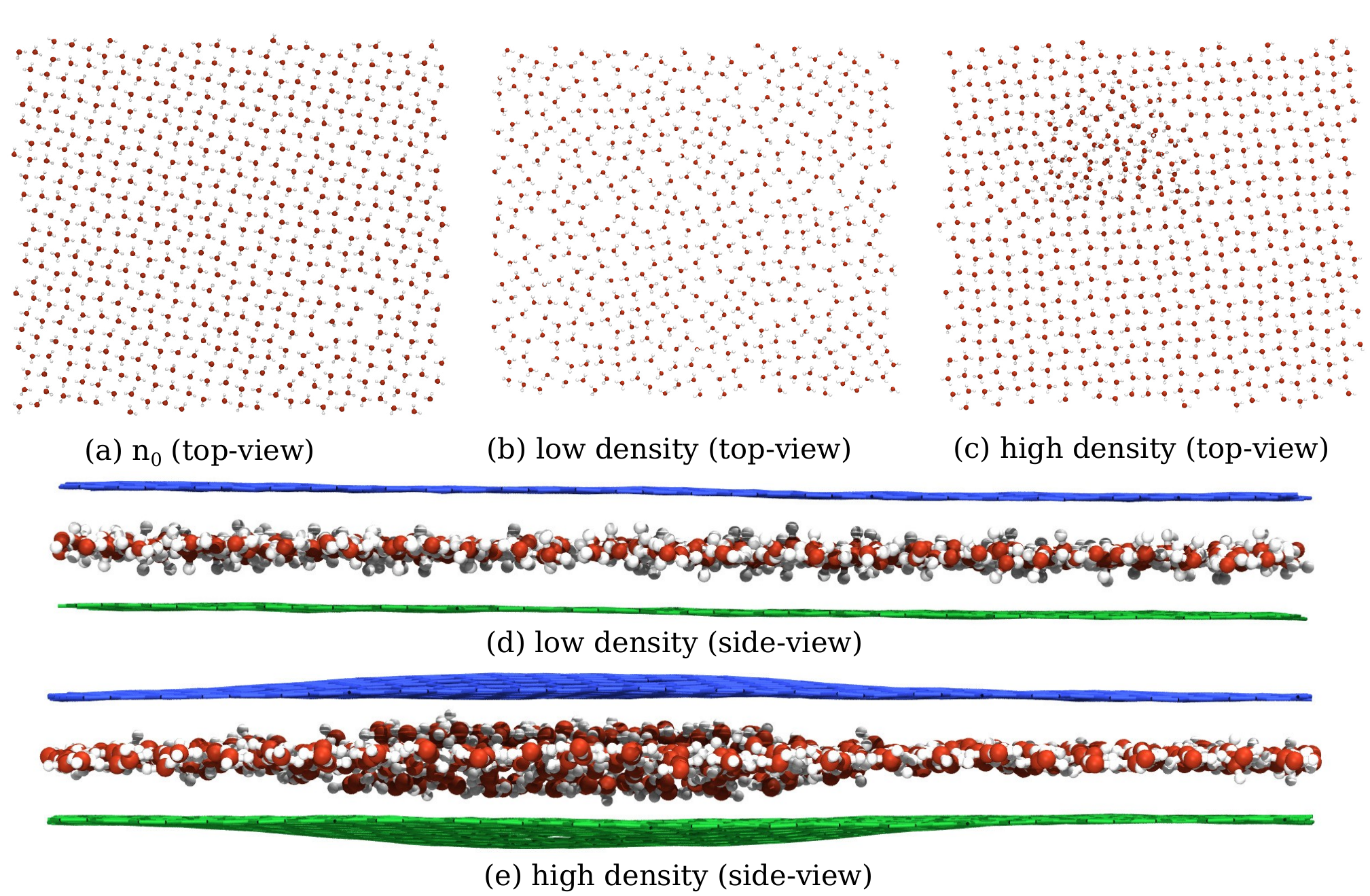}
    \caption{
    Snapshot of water structure in case of (a) natural filling $n_0$, (b) low density filling $n_V$, and (c) high density filling $n_D$ at twist angle $\SI{16.4}{\degree}$. 
    The side view of the water structure in case of low density filling (d) contrasts sharply with the side view of high density filling (e) where a dripplon formation takes place and distorts the slab geometry of the confined region. 
    The low density filling can be characterized by the presence of voids and imperfect square ordering at some places in the water structure as shown in (b).
    }
    \label{fig:density_dependent_structure.pdf}
\end{figure}

Structure and dielectric properties discussed in the previous sections are also calculated for low density and high density at the five aforementioned angles, and the results are plotted on the same figures along with $n_0$ (Fig.~\ref{fig:twist_ktic_50ns_inset.pdf}, \ref{fig:twist_epsilonAvgPerp_50ns.pdf}).
A low or high density than the natural filling ($n_0$) gives rise to lower order parameter value due to formation of "dripplon" \cite{yoshida2018dripplons} and voids/defects in the structure respectively.
In Fig.~\ref{fig:density_dependent_structure.pdf} we show this dripplon and void formation from our simulation snapshots at twist angle $\SI{16.4}{\degree}$, for all the three cases - $n_0$, low and high density.
In low density filling, patches of improper square ordering and void spaces are there along with ordered region at a few locations, which constantly changes with time. 
Similarly, for high density filling, the region of dripplon formation shown in Fig.~\ref{fig:density_dependent_structure.pdf}, gives water molecules of that region more freedom to contribute in perpendicular static permittivity than the remaining square ordered structure. 
However, the dripplon increases the channel width and a result of which the volume of confinement increases.
This consequently compensates for the higher fluctuations in the dripplon region, and ultimately results in a value of dielectric constant that is 0.3 lower than natural filling and low density filling.

\begin{figure}[ht]
    \centering
    \includegraphics[width=0.8\textwidth]{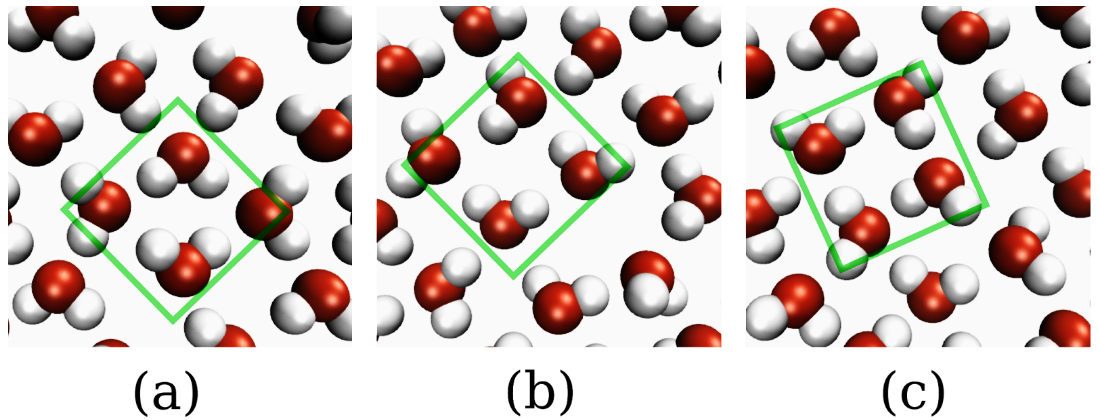}
    \caption{
    Snapshots of confined TIP4P/2005 water in graphene nano-channel showing the formation of square ordering for $\SI{12.5}{\per \nano \meter \squared}$ density filling. (a) $\SI{0.0}{\degree}$ twist  (b)$\SI{16.4}{\degree}$ twist, (c) $\SI{29.4}{\degree}$ twist. A green box is shown in each case to identify one such square arrangement in the monolayer-ordered structure.
    }
    \label{fig:structure_tip4p2005_density_12.5.png}
\end{figure}

\begin{figure}[ht]
    \centering
    \includegraphics[width=0.5\textwidth]{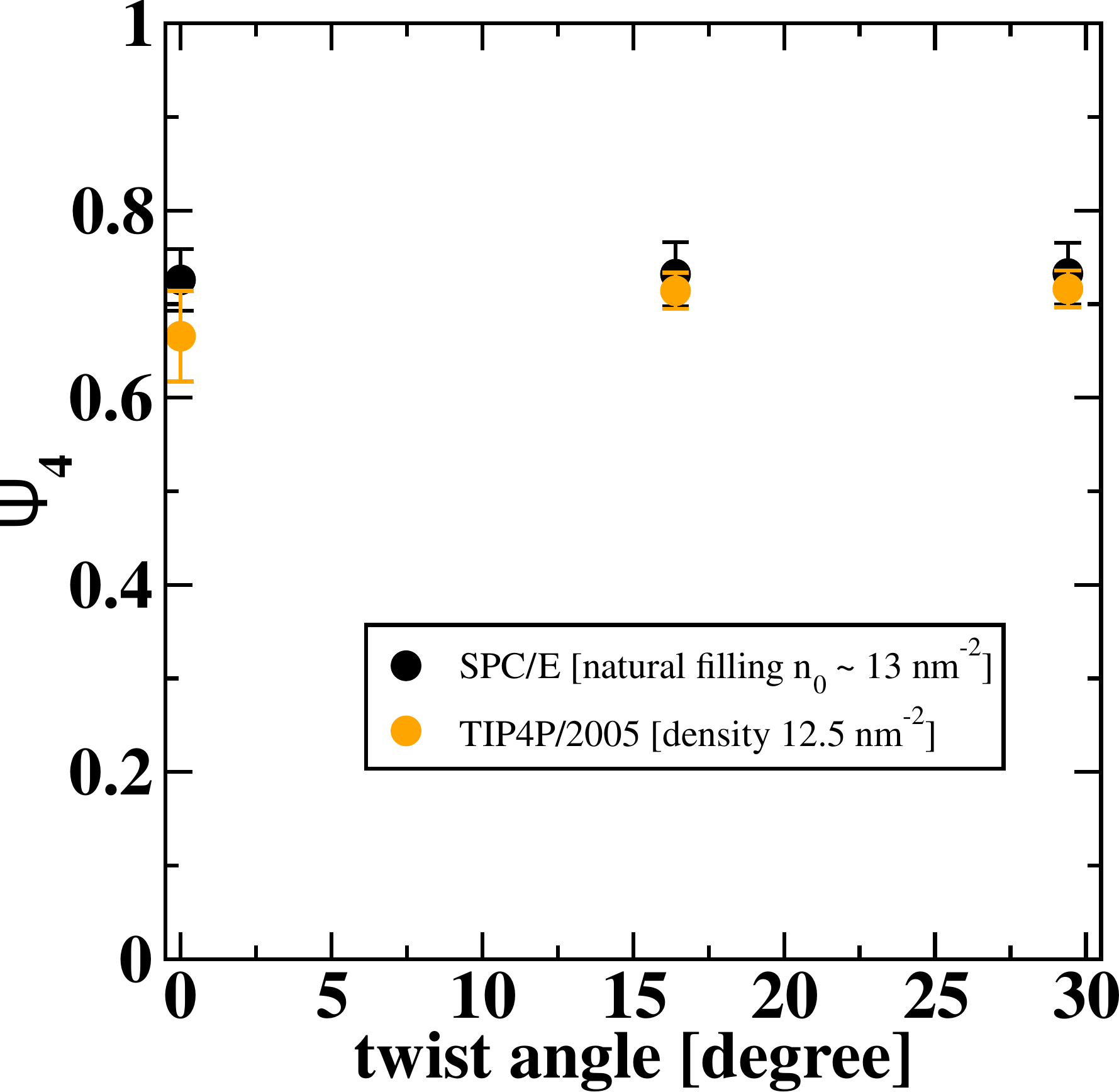}
    \caption{
    Order parameter value of confined TIP4P/2005 water in graphene nanochannel with a density of $\SI{12.5}{\per \nano \meter \squared}$. For a clear comparison with SPC/E model, we have also marked its corresponding values at those twists.  The order parameter is lower than the corresponding value of SPC/E because there are some voids, and overall the structure of SPC/E is stable forming square ordering even at lower density unlike TIP4P2005.
    }
    \label{fig:psi4_tip4p2005_density_12.5.pdf}
\end{figure}

For TIP4P/2005 water, we have filled the channel with densities of \SI{12}{\per \square \nano \meter}, \SI{12.5}{\per \square \nano \meter} and \SI{13}{\per \square \nano \meter}. 
We find that at \SI{12.5}{\per \square \nano \meter} density, TIP4P/2005 starts showing squared-ice ordering just like it was observed for SPC/E in its natural density filling.
Figure \ref{fig:structure_tip4p2005_density_12.5.png} shows snapshots and Fig.~\ref{fig:psi4_tip4p2005_density_12.5.pdf} shows the $\Psi_4$ value at three twist angles explored to investigate this phenomenon. 
For comparison, the value of SPC/E is also plotted on the same figure. 
We have compared this critical density (\SI{12.5}{\per \square \nano \meter}) of TIP4P/2005 with our data from our earlier work where we had obtained similar water structures under high-pressure conditions inside two rigidly held finite graphene sheets.
Analysis shows the high-pressure simulations gave rise to similar density inside the channel when square-ice ordering was obtained in the case of TIP4P/2005. 

\section{IV. Discussions and conclusion}
\label{discussion_and_conclusion}
To summarize this work, we have explored the dependence of confined water structure, dielectric and frictional property with small angle twists of bilayer graphene using classical MD simulations.
The structure of confined water is found to be insensitive to the twist angle of the tBLG and all the equilibriated structure shows a square-ice ordering as reported in experiment by Algara \emph{et. al.} \cite{algara2015observation}. 
The angle distribution of water arrangement and the radial distribution function shows almost indistinguishable effect of twist, which agrees with our previous observation on untwisted graphetic confinement \cite{majumdar2021dielectric}.

All the twist angles show anomalously low permittivity of confined water as were observed in experiments and many theoretical works.
And also through this work we have shown that this static permittivity of confined water have negligible variation with twist angle difference of tBLG.

We further analysed the friction coefficient of this confined water and we found that the correlation convergence time is significantly more for smaller twist angles.
We have found that friction is dependent on twist angle for small angles below $\SI{5.1}{\degree}$ and becomes insensitive to twist for higher angles.
Additionally, we carried out simulations at five twist angles with chosen high and low density of water inside infinite tBLG, and found that both these cases show reduction in overall ordering and increased dipole fluctuation from the natural filling density.
The results indicate that although density also plays a crucial factor with very high sensitivity.
From an experimental perspective, these results ease the overall burden of fine tuning for small twist angles for controlling monolayer confined water in graphene channels from an electronic device perspective. 
However, the small twist angles may be of significant importance in the context of wetting and desalination based applications. 

\begin{acknowledgement}

We thank DST and MHRD, India for financial support. 
We thank TUE-CMS, IISc for providing the computational facilities

\end{acknowledgement}

\bibliography{reference}

\end{document}